\documentclass[a4paper,11pt]{article}
\usepackage{jinstpub} 
\usepackage{lineno}
\usepackage{comment}

\title{\boldmath A scanning device for spatial quantum efficiency measurements of photomultipliers tubes}
\author[a]{P. Migliozzi,}
\author[a]{C.M. Mollo,}
\author[a]{A. Simonelli}

\affiliation[a]{INFN, Sezione di Napoli, Complesso Universitario di Monte S. Angelo, \\ Via Cintia ed. G, Napoli, 80126 Italy  \\
}

\emailAdd{andreino.simonelli@na.infn.it}

\abstract{This paper presents a quantum efficiency measurement setup based on a 2D motorized stage, a wide spectrum xenon lamp, a beam splitter system, and two calibrated photo-diodes for measuring the quantum efficiency (QE) of photomultiplier tubes (1 to 10 inches).
We will demonstrate the effectiveness of technical refinements on the measurements procedures over some existing setups already shown in literature.
The large area covered by the 2D stages permit to study the quantum efficiency of PMTs with diameter up to ten inches.
The results obtained will show the high precision and accuracy in characterizing the quantum efficiency versus wavelength over the range of 250 nm to 1100 nm and along the photo-catode surface. The setup monitors the light intensity synchronously with the output current yield from photosensors under test. This ensures the accuracy and repeatability of the measurements. The motorized stage allows precise positioning of the light source with respect to the active area. The emission spectrum of the xenon lamp provides a broad range of illumination in terms of dynamics and wavelength span.}

\keywords{photomultipliers, quantum efficiency}
\begin{document}
\maketitle
\flushbottom
\section{Introduction}
\label{sec:intro}
In the realm of astroparticle physics, where the study of high-energy cosmic particles through the detection of Cherenkov radiation, the dependable performance of photomultiplier Tubes (PMTs) stands as a critical factor \cite{aharonian2006hess, aleksic2016magic, aartsen2017icecube}. These experiments rely on the extraordinary sensitivity of PMTs to low-intensity light, particularly Cherenkov radiation emitted when charged particles surpass the local speed of light. The pivotal role of Cherenkov light extends to investigating high-energy cosmic rays, gamma rays, and neutrinos. The QE of PMTs, representing the likelihood of detecting photons at a specific wavelength, is foundational for accurately discerning the energy, direction, and characteristics of incident particles. In diverse environments like Earth's atmosphere, underground locales, or underwater setups, precise QE measurements become indispensable. This paper emphasizes the critical role of a stable and user-friendly setup for accurate QE measurements for PMTs in astroparticle physics experiments. Such measurements directly influence the sensitivity and reliability of Cherenkov telescopes and detectors, providing reliable data for Monte Carlo simulations. A proficient grasp of QE measurements is integral to advancing our comprehension of high-energy cosmic phenomena. This paper aims to shed light on the indispensable role played by PMTs in unraveling the enigmas concealed within the high-energy realms of the universe, utilizing the lens of QE measurements. Furthermore, the spatial variations of QE across the photo-cathode surface constitute a critical parameter for PMTs, serving as a crucial input for Monte Carlo simulations and data analysis. Consequently, our focus is directed towards exploring the limits in resolution within the scanning capabilities of a newly developed motorizer setup.
The acquisition of a substantial data set of measured PMTs is paramount for conferring statistical significance to variables such as QE. Hence, complete automation of this process is imperative, facilitating the sampling of a large set of specimens.
Our proposed setup, featuring a 2D motorized stage, a wide-spectrum Xenon lamp, a beam splitter system, and two calibrated photo-diodes, is expressly designed for PMTs. Notably, its adaptability allows for seamless extension to other photosensors with minimal modifications. The operational simplicity of our system is underscored by the ease of use – a mere push of a button and placement of the sample in the dedicated slot initiates the process.
\section{State of the art}
The quantum efficiency is the ratio of the number of photoelectrons emitted by the photo-cathode to the number of photons incident on it. The effect was predicted by \cite{Einstein1905} in 1905 and still today the impact of this fundamental study is huge from the everyday life to the latest applications for research in fundamental physics. This setup is aimed at the characterization of photomultiplier tubes who's sensitivity is ranging from 250 to 1100 nm. QE is a crucial parameter for evaluating the performance of photo-sensor devices. It measures the proportion of incident photons that are converted into electrical charge carriers. Two primary techniques are employed for QE measurement: the absolute radiometric technique and the correlated photons technique. 
The correlated photons technique \cite{padgett2019,kwiat1994,migdall1995} directly measures the number of incident photons and the number of generated electrons, providing an absolute QE measurement. This technique is based on parametric down-conversion (PDC), where a high-energy photon is split into two lower-energy photons. One photon illuminates the photo-sensor device, while the other triggers a detector. The QE is calculated from the coincidence rate between the two detectors. 
On the other hand the absolute radiometric method is based on an experimental procedure that involves the utilization of a calibrated light source possessing a precisely known spectral power distribution. This light source, covering the relevant wavelength range for the photo-detector, serves as the means to illuminate the detector. Subsequent to the precise measurement of the incident optical power using a calibrated power meter or radiometer, the photo-detector's response in terms of output current is meticulously recorded. The quantum efficiency is then computed as the ratio, thus providing a direct and quantitative characterization of the photo-detector's performance $QE(\lambda) = \frac{I}{P_{\text{in}}(\lambda)}$ where I is the phtocurrent and $P_{\text{in}}(\lambda)$ is the wavelenght dependent power that enlightens the photo-sensor . Critical considerations within this method encompass the necessity for meticulous alignment and calibration procedures, ensuring the accuracy and traceability of measurements. Absolute radiometric QE measurements offer a straightforward and cost-effective approach for determining this crucial parameter. However, the requirement for a calibrated reference photo-diode limits the absolute accuracy of these measurements. In this paper we try to push the limits of this method. The method can have different experimental interpretations and here we will examine some papers where the results are described specially on PMT tubes and finally compare with our setup highlighting the improvements
In \cite{Bueno2008} the setup consists of a Xenon lamp, a computer-controlled monochromator, and a light-tight box with a reference photo-diode and the tested PMT. The lamp, stable with deviations below 1$\%$, is pulsed, covering 300 nm to 600 nm. The monochromator allows precise wavelength selection. A 50/50 polka-dot beam splitter and calibrated components ensure accuracy. Independent measurements using a pico-ammeter and an electrometer determine quantum efficiency by comparing currents from the reference photo-diode and PMT first dynode. A 300 V gradient optimizes collection efficiency. Systematic effects from the beam splitter are addressed by exchanging positions in two measurements, and the geometric average provides the final quantum efficiency value.
In \cite{MIRZOYAN2006230} to achieve their measurements, they illuminated the central 15 mm diameter of PMT photo-cathodes and measured the photocurrent after applying a 200V potential difference between the photo-cathode and the 1st dynode. A calibrated 10mm×10mm diode from Hamamatsu served as the reference detector in all measurements. A spectrophotometer, controlled by Lab-View under computer control, was assembled using commercial parts along with the calibrated diode. They calculated the error of the QE measurements to be 2.34$\%$, with the major contribution coming from the 2$\%$ calibration precision of the used photo-diode.
The quantum efficiency at cryogenic and ambient temperatures is measured by \cite{Lyashenko_2014} using a setup based on absolute radiometric method. The measurement setup involved comparing test photomultiplier Tube (PMT) photo-currents with a National Institute of Standards and Technology (NIST)-calibrated reference photo-diode (PD). The PMT chamber and spectrometer, separated by an MgF2 window, operated under high and low vacuum conditions, respectively. Two light sources (deuterium and tungsten lamps) covered spectral ranges from 154.5 nm to 400 nm and 300 nm to 600 nm. Currents from the reference and test PMTs, along with the PD, were read using picoammeters. The PMTs operated with a 300 V positive bias at the first dynode in DC current mode. Measurements involved varying the spectrometer wavelength, normalizing currents to the reference PMT, and averaging over 50 measurements. A 100 msec integration time for each current reading ensured a 5-second measurement time at each wavelength. The positive 300 V bias guaranteed nominal photoelectron collection efficiency, normalized by the reference PMT current to compensate for light fluctuations. The entire process was automated using LabView software.
\cite{MIRZOYAN2007449} has studied ultra bi-alkali PMTs, in their measurements, a custom-assembled spectrophotometer using readily available components was employed. For reference, a Hamamatsu calibrated diode with dimensions of 10×10 mm, boasting a calibration precision of 2$\%$, was utilized. Both the spectrophotometer and the diode were controlled and monitored through LabVIEW under computer supervision. The central area of the PMT photo-cathode, ranging from 5 to 15 mm in diameter, was typically illuminated. A potential difference of 200V was consistently applied between the photo-cathode and the first dynode, followed by the measurement of photocurrent using a Keithley 485 picoammeter, with all other dynodes shorted with the first one. 
In the context of Dama Libra experiment \cite{RBernabei_2012} measurements were carried out on PMT tubes utilizing laser sources (Ar+, He-Ne, Nd:YAG, and laser diodes) in the 351–1500 nm range. Laser intensities were measured with power meters and calibrated filters for low values. PMTs were electrically connected with the anode and all dynodes at +200 V, while the photo-cathode was grounded with a resistor. The Keithley electrometer mod. 6514 measured cathode current (IK). This work reports also a radial scan of the QE.
This overview of the setups documented in literature allow us to highlight some novelty elements for our setup like: the 2d motorized stage, a state of the art calibrated couple of photo-diodes with the highest accuracy on the market, a wider range of wavelengths available for device under test QE, the use of a feedback loop stabilized source ensuring the highest possible stability for a lamp source, a six decades optical power dynamics.

\section{Setup details}
\begin{figure}[h!]
\centering
\includegraphics[width=.7\textwidth]{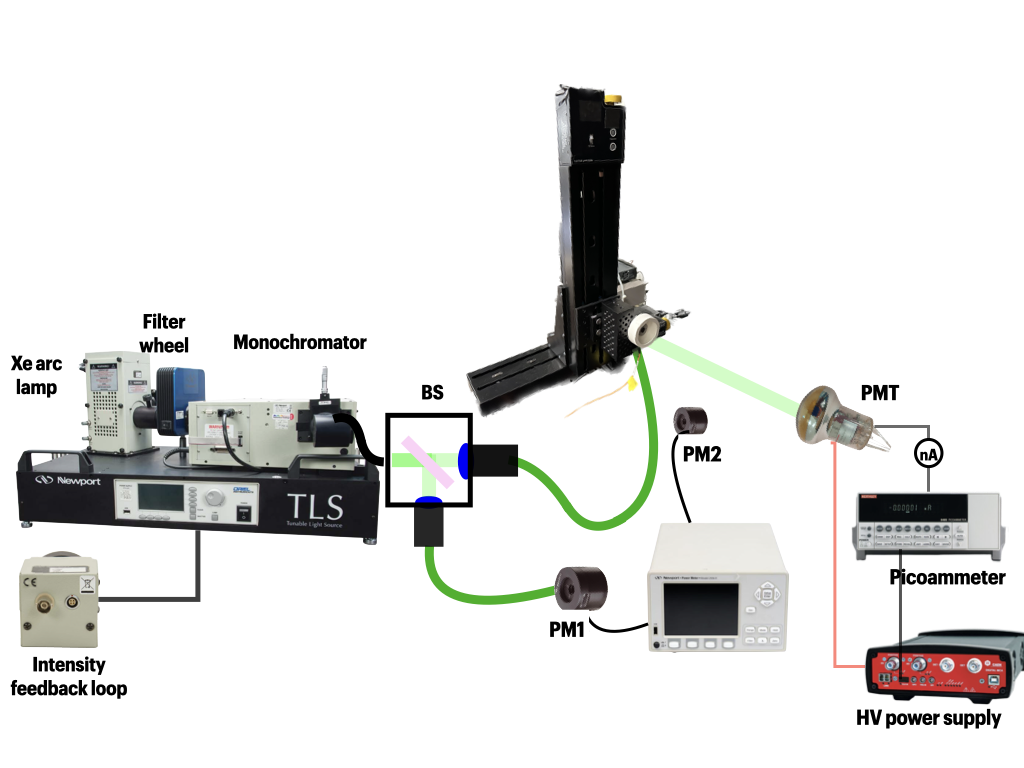}
\caption{A not to scale sketch of the setup and the optical and electrical connections }
\label{SCHEMATIC_SETUP}
\end{figure}

\begin{figure}[h!]
    \centering
    \includegraphics[width=.99\textwidth]{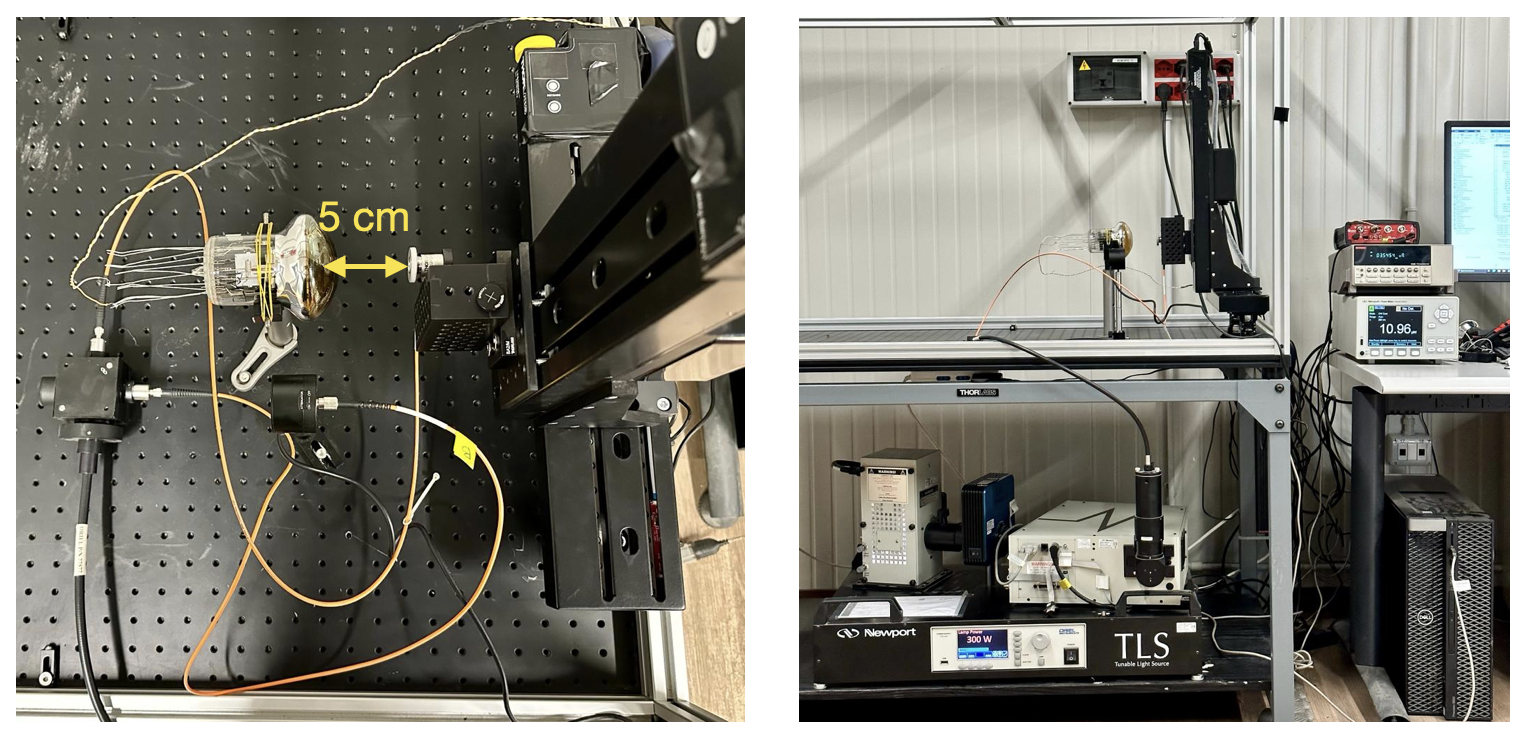}
    \caption{On the left picture a detail over the optical bench setup showing the beam splitter and the device under test and the calibrated power meter, on the right picture the ensemble of the instruments}
    \label{picofsetup}
\end{figure}

\begin{figure}
    \centering
    \includegraphics[width=.95\textwidth]{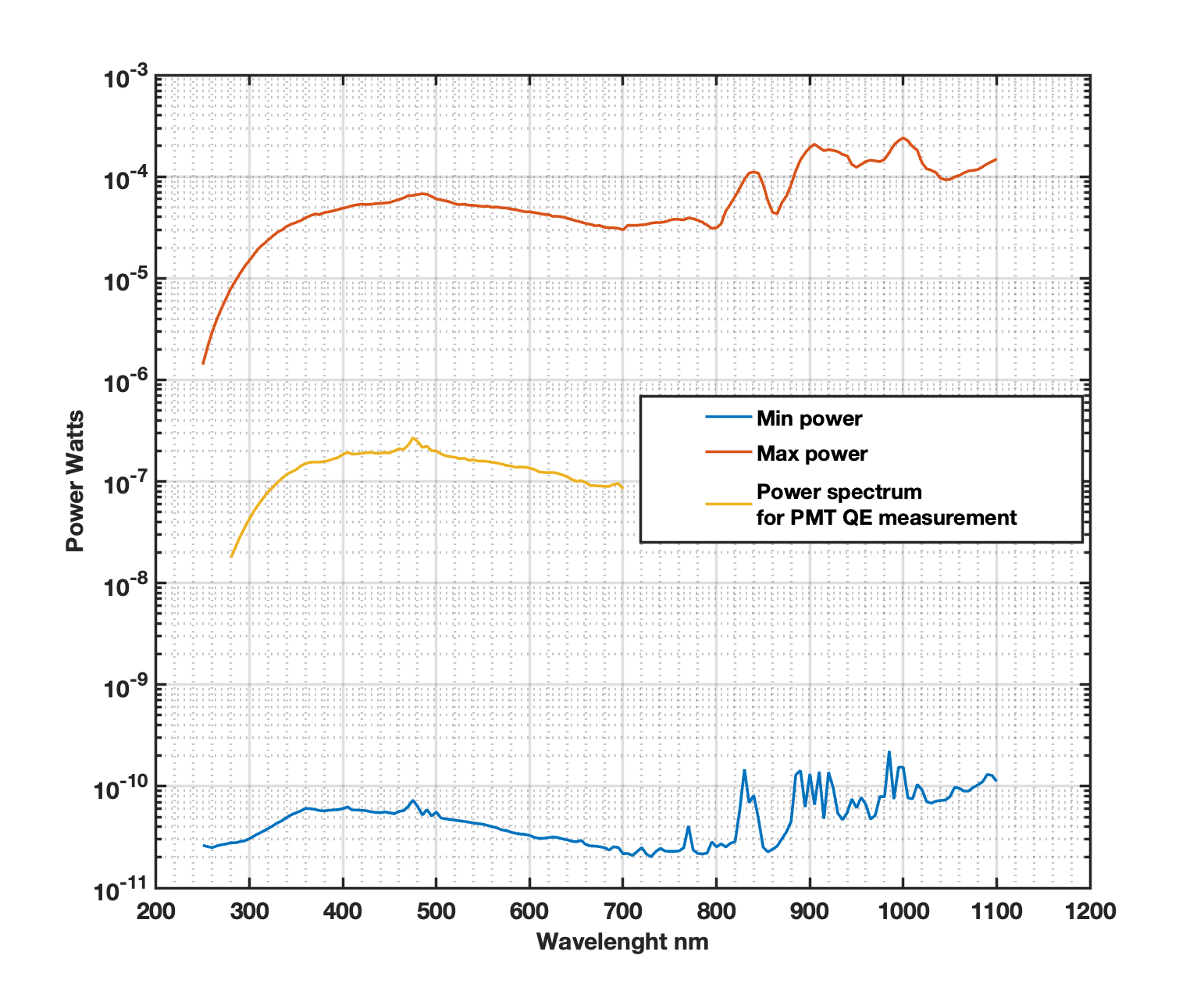}
    \caption{Maximum and minimum power deliverable to the device under test and typical power used for quantum efficiency measurements on the first dynode}
    \label{DUT_PWR}
\end{figure}

The setup shown in figure \ref{SCHEMATIC_SETUP}  allows us to characterize quantum efficiency versus wavelength and make a map of QE over the surface of photo-cathodes with high precision and accuracy. All the components and instruments composing the device are depicted in a schematic way for better comprehension of the connections and light paths in figure \ref{picofsetup}. To estimate the quantum efficiency, two calibrated power probe (Newport 918D-UV-OD3R) are used to measure the power of the incident light and the splitting ratio of the beam splitter outputs. This calibrated power probe are connected and read by a Newport 2936 R base. This ensemble of photo-diode and detector base represents the state of the art of commercial power meter ensuring a NIST traceable calibration and the highest accuracy on the market. The calibrated NIST photo-diode exhibits varying uncertainties across different wavelength ranges: 3.4$\%$ uncertainty in the 220-300 nm range, 1.65$\%$ in the 300-430 nm range, 1.1$\%$ in the 430-1000 nm range, and 4.3$\%$ in the 1035-1065 nm range. A picoammeter (Keithley 6485) is used to measure the current obtained on the first dynode. The use of this instrument is reported also in other works and if properly maintained can ensure a 0.4$\%$ accuracy error in the 20 nA range which is the one in use for PMT QE measurements. Useless to say that the calibration of this instruments comes on top of all the experimental procedure described below, but these are the foundations of this calibration dependent method. Provided this, there are many experimental solutions to obtain the QE values. In this work the authors payed a lot of attention to try to reduce at minimum the uncertainties and enhance the user friendliness of the setup. This in order to allow massive measurements on hundreds of PMTs that can be performed by trained technicians. The output of the photomultiplier represents the convolution between the xenon lamp spectrum and the detector response. To prevent underestimation of the emitted photocurrent, the photo-cathode is maintained at a voltage of -300 V with respect to the first dynode, which is grounded along with the remaining nine dynodes and the PMT anode. This approach maximizes photoelectrons capture, i.e. those ones that may escape from the first dynode collection. 
The light used in the experiment is generated by a Newport TLS 260 tunable light source, which is a 300-watt xenon arc lamp ensemble known for its high UV output and excellent spectral source. 
The range of available power at the device under test is shown in figure \ref{DUT_PWR} where the range between the maximum and minimum measurable power by our system is seven orders of magnitude allowing future expansions and developments to different kind of photo-sensing devices.
The light emitted by the xenon lamp is passing trough a filter wheel motorized which is equipped with four low pass filters. This is meant to avoid higher order diffraction which translates in stray-light and in the end systematic in QE measurements. This phenomenon is prevented with the filter wheel but is also intrinsically minimized by the Czerny-turner monochromator geometry that ensures the best spectral purity among other monochromators design.
The light exiting the monochromator is coupled to a high grade fused silica fiber bundle which transports the light to the beam splitter cage.
The light is splitted by a polka dot beam splitter with mirror dots spacing of 56 $\mu$m. The choice of this beam splitter was driven by several considerations like the constant reflection to transmission ratio over range of interest, a good approximation to 50:50 splitting ratio and the relatively insensitivity of metal coating to incidence angle.
The minimum spot size over the beam splitter to achieve a good 50:50 ratio is 2 mm. This is easily achievable given the high numerical aperture of the fiber bundle, but poses another problem i.e. the collimation of the light in a second step after the transmission and reflection of the splitted beams. This is achieved by customizing with a UV enhanced fused silica lenses multi-mode collimator. This is fundamental to do not cut out the precious information of QE below 300 nm. 
The light is then transmitted via two multi-mode 600 $\mu$m or 200 $\mu$m core (the core diameter is chosen according to desired spot size on the PMT cathode) 1,5 meters long  optical  fibers. The optical fibers are made of Pure Silica core and fluorine-doped silica cladding with intrinsic very low auto fluorescence.
A possible source of stray-light can be the self fluorescence of optical fibers when exposed to UV light. This light generated inside the fiber can give false results when measuring the QE specially in the UV region. The high purity fused silica and the cladding quality minimizes. This phenomenon that was also experimentally proven to be negligible by checking the UV-VIS spectrum of the transmitted light from the optical fiber with a CCD spectrometer. 
The choice of the core depends on the needed attenuation and spot size.
One output of the beam splitter goes to the power meter and the other goes to a parabolic mirror collimator RC02FC-F01-UV-enhanced from Thorlabs .
\begin{figure}[h!]
\centering
\includegraphics[width=.9\textwidth]{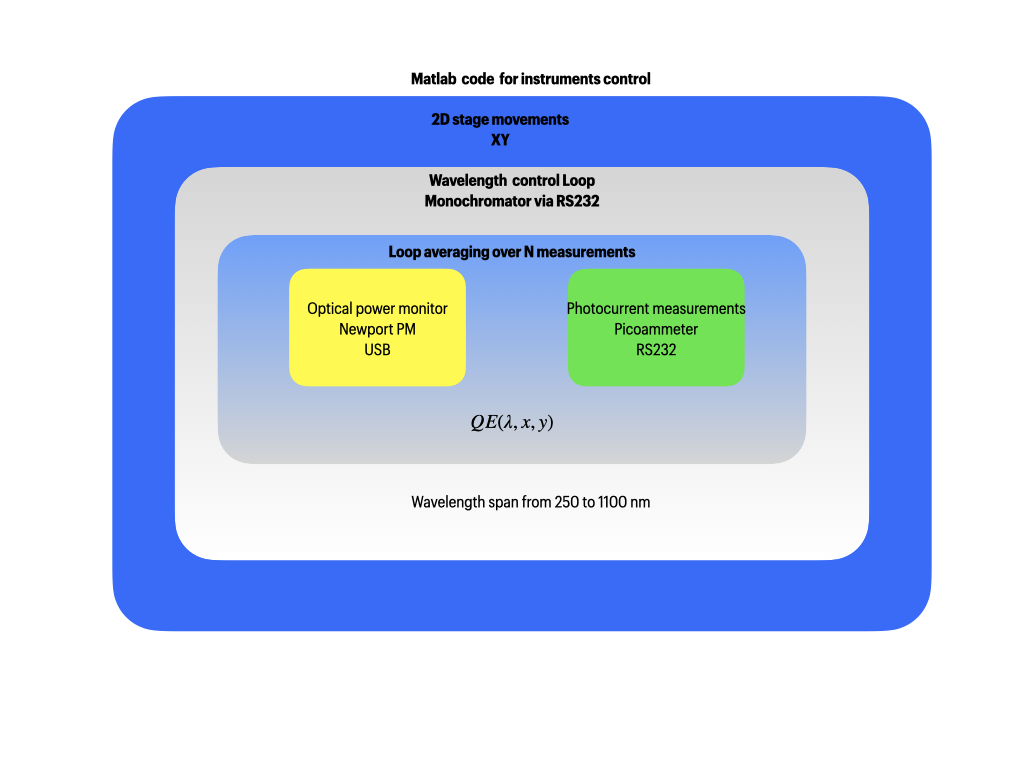}
\caption{Scheme of the control software\label{scheme}}
\end{figure}
The advantage of using a parabolic mirror collimator is that it maintains a constant focal length at different wavelengths, ensuring a consistent spot diameter for testing the photo-catode. Additionally, it solves the issue of having a 90-degree output with respect to the fiber connector in this specific mechanical assembly. To ensure the stability of the light intensity, a feedback loop is used. This involves using a thermo-electrically cooled silicon photo-diode monitoring the broad band lamp emission. The signal from the feedback photo-diode is sent to the feedback loop, which controls the power of the lamp to stabilize the light intensity.
\begin{figure}[htbp]
\centering
\includegraphics[width=.68\textwidth]{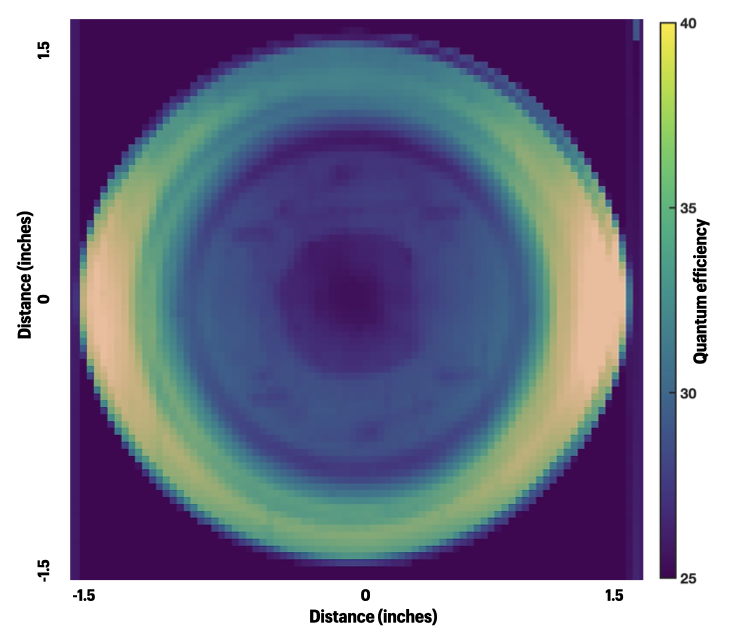}
\qquad
\includegraphics[width=.57\textwidth]{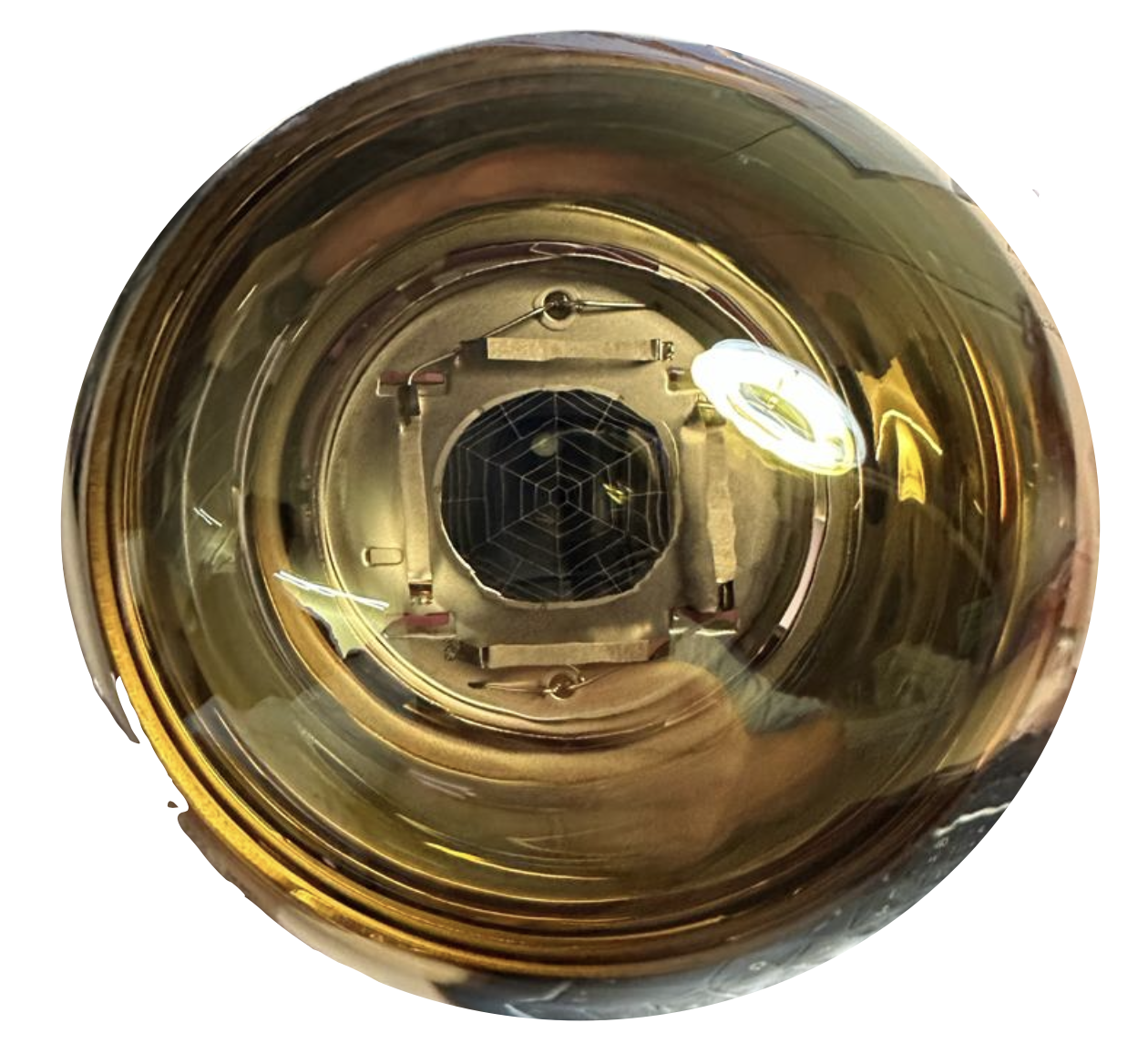}
\caption{A comparison between the two dimensional scan of the photo-catode at 410 nm (top) and a photography of the cathode itself (bottom) enlightening the scanning capabilities of the setup}
\label{2D_qe}
\end{figure}
By using this feedback loop, the light intensity can be maintained at a constant level, which is important for ensuring the repeatability of the measurements improving precision as shown e.g. in \cite{LaserFocusWorld}. This active stabilization via feedback together with the simultaneous measurement of photo-current and power permits to reduce to minimum the fluctuation of the observable QE.
\begin{figure}[h!]
    \centering
    \includegraphics[width=1.0\textwidth]{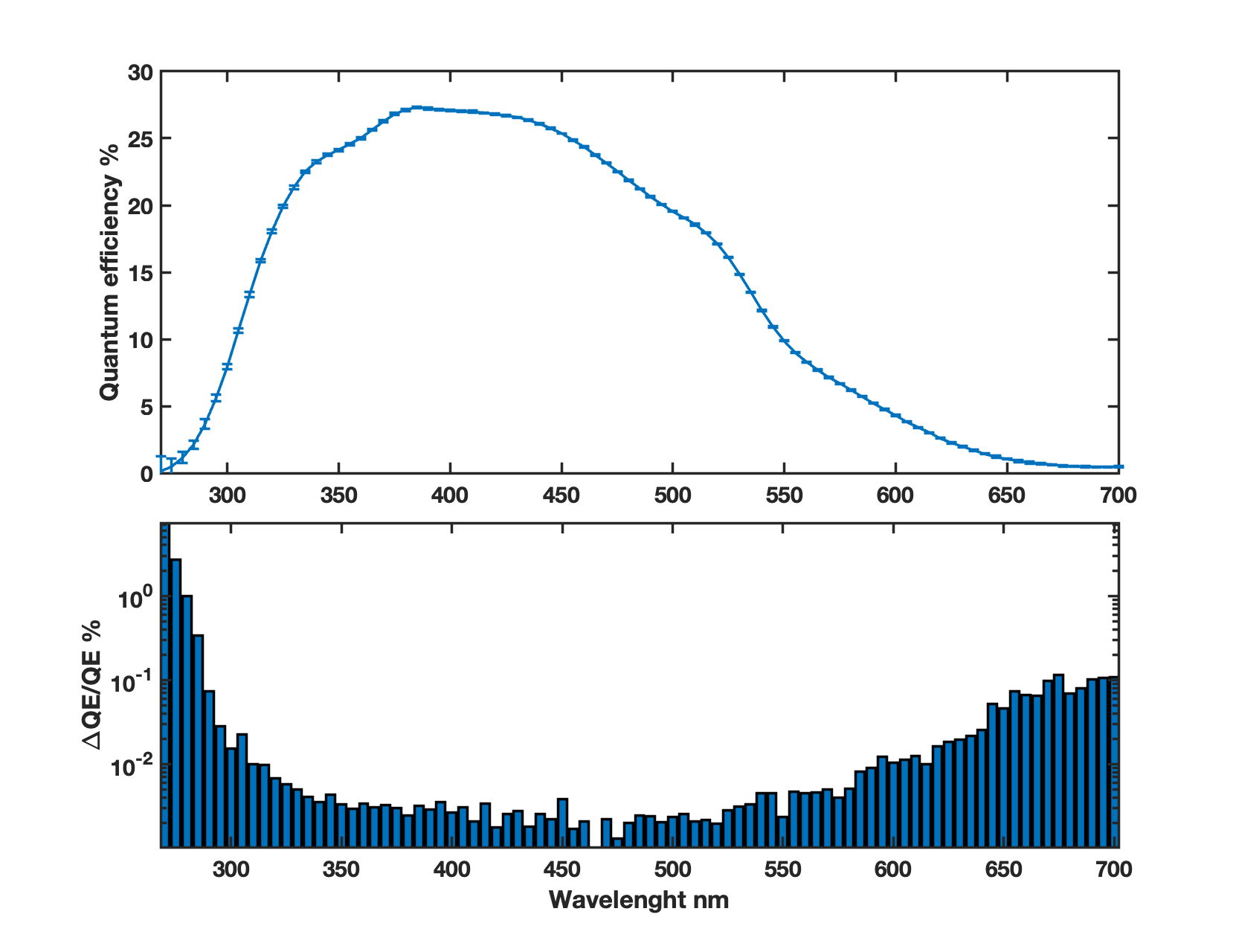}
    \caption{(Top) Quantum efficiency versus wavelength for R14374 Hamamatsu PMT. (Bottom) Relative error for this specific QE acquisition}
    \label{laqe}
\end{figure}
This is particularly important when taking multiple measurements at different wavelengths, as variations in light intensity can affect the results. 
\begin{figure}[h!]
    \centering
    \includegraphics[width=0.99\textwidth]{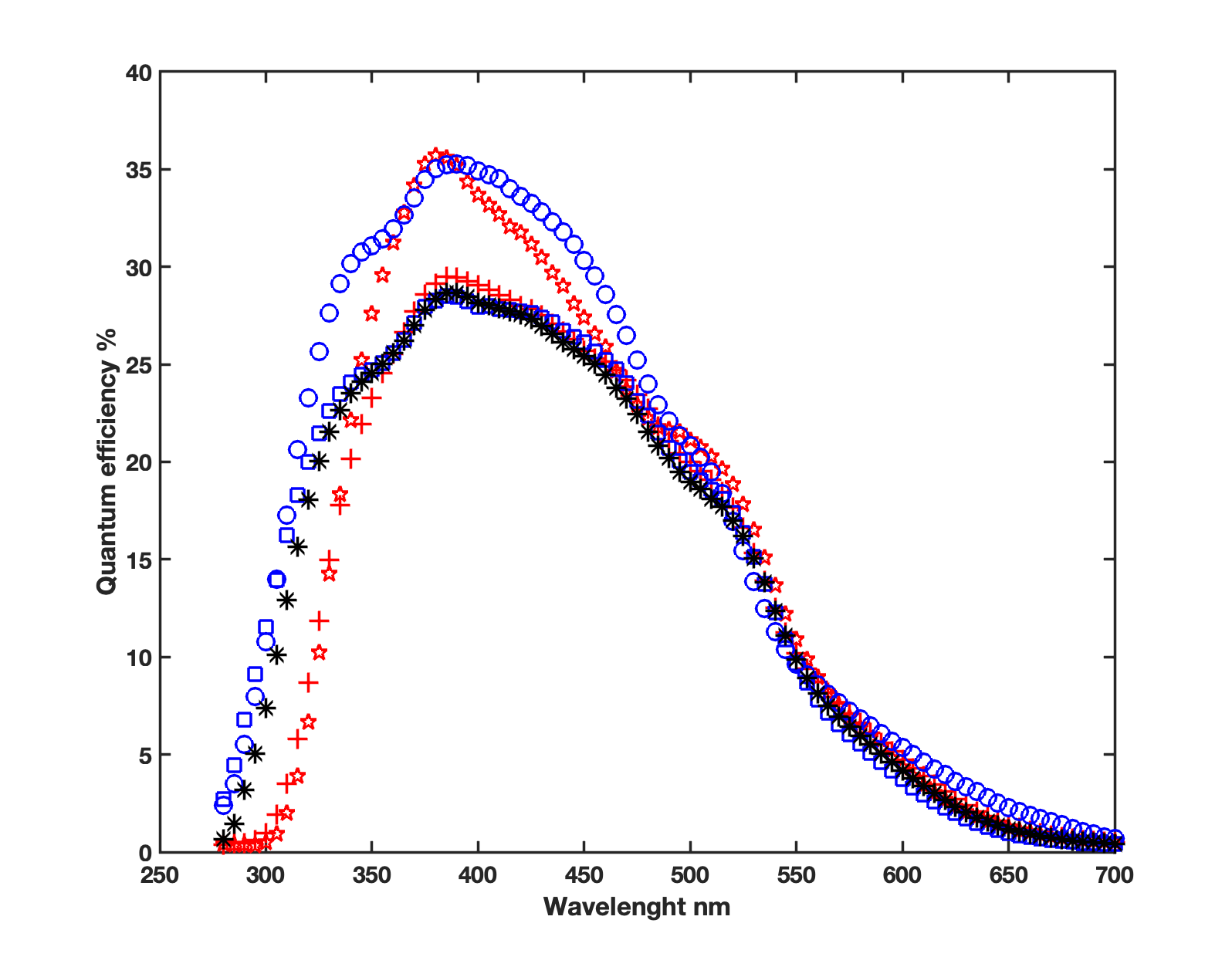}
    \caption{Comparison of five available bi-alkali metal 3 inch PMTs}
    \label{laqecompare}
\end{figure}
Finally the light collimator is mounted on the Z axis dedicated stage which is mechanically coupled to the X axis stage. These stages are LTS300C from Thorabs, this versatile system features a 300 mm travel range. With a load capacity of 15 kg (33.1 lbs) when horizontally mounted and 4 kg (8.8 lbs) when vertically mounted, it accommodates various setups. The system boasts a maximum velocity of 50 mm/s, ensuring swift and precise movements, and offers bidirectional repeatability of <±2 µm and accuracy of 5 µm. This is far below the actual spot size used for cathode scanning offering future improvements in scanning resolution.
The Matlab control code is based on the instrument control toolbox and the general concept behind this code is shown in figure \ref{scheme}. 
The communication with the peripherals and instruments is obtained via RS-232 in the case of the Keithley picoammeter and of the monochromator. The power meter is controlled via USB by means of a dedicated library available at \cite{MathWorksFileExchange}.
Finally the 2D stage is controlled using drivers developed in \cite{JulianFells} and adapted for our scopes.
The synergy of all the instruments as well as the synchrony permits large customization freedom for the code as well as for the setup.
The inner loop of figure \ref{scheme} averages the simultaneous power and current measurements in a user settable time interval. The outer loop repeats the inner loop measurement in a user definable wavelength set of values that can be scattered or incremental according to the sampling needs. Finally the outermost loop allow the control of the movement of the 2D stage in order to map the active area of a photo-sensor. In our case the photo-catode of a 3 inches PMT.  By disabling the outermost loop and choosing a fixed position of the light insisting on the active area a wavelenght dependent at fixed position QE can be obtained. 
\section{Results}
We present the outcomes derived from the spatial mapping of the photo-cathode of a Hamamatsu R14374 bi-alkali 3-inch PMT. The mapping was conducted at a fixed wavelength of 410 nm with a step size of 500 $\mu$m. The scanning protocol follows a snaking path across the PMT area. Point-wise measurements were acquired by averaging ten sets of optical power and photo-current measurements, yielding the Quantum Efficiency (QE) pixel, as illustrated in figure \ref{2D_qe}. Remarkably, the obtained map offers intricate details of the photo-cathode, comparable to a photographic representation, revealing internal metal structures.This result underscores the remarkable resolution capabilities of the setup, simultaneously prompting scientific inquiries regarding the conceptualization and definition of quantum efficiency. The visibility of inner photomultiplier structures is attributed to their reflective properties. Photons not contributing to the photoelectric effect in the coating during the initial pass are reflected or diffused by the metal components. Some of these photons, upon reflection, induce secondary photoelectron emission. The QE as a function of wavelength for the same PMT was measured at the center of the photo-cathode, employing the same strategy as the two-dimensional scan. The wavelength step size is 5 nm, and the result is depicted in figure \ref{laqe}. The reported errors in the QE curve are derived as the standard deviation from ten measurements per wavelength. The lower section of figure \ref{laqe} presents a bar plot illustrating this reproducible error, deemed as accuracy. In addition to this, an accuracy-driven error must be considered, arising from calibration uncertainties in the power meter and picoammeter. This can be estimated using error propagation, as outlined in Table \ref{errors}. 
The overarching result of error budget estimation reveals a notable constraint: the limiting factor is, in fact, the accuracy of the instruments, as evidenced by the relative errors of repeatability illustrated in the lower plot of figure \ref{laqe} and the accuracy error attributed to instrument calibration detailed in Table \ref{errors}. This implies that the current measurement strategy is poised for enhancements in instrumentation, paving the way for a more precise instrument for Quantum Efficiency (QE) testing in the future.
Further test measurements were conducted and are illustrated in figure \ref{laqecompare} for four different three-inch PMTs from various brands with distinct efficiencies. For clarity in graphical representation, error bars are omitted. The final plot demonstrates the capability to discern quantum efficiency characteristics among different commercially available PMTs with varied brands and coating technologies.
\begin{table}[h]
    \centering
    \begin{tabular}{cccc}
        \hline
        \textbf{Wavelength} & \textbf{Power meter accuracy(\%)} & \textbf{Picoammeter accuracy (\%)} & \textbf{Total Error (\%)} \\
        \hline
        220-300 nm & 3.45 & 0.4 & 3.52 \\
        300-430 nm & 1.67 & 0.4 & 1.71 \\
        430-1000 nm & 1.13 & 0.4 & 1.16 \\
        1035-1065 nm & 4.33 & 0.4 & 4.35 \\
        \hline
    \end{tabular}
    \caption{Instrumental error for each wavelength range}
    \label{errors}
\end{table}
\section{Conclusions}
The scope of this paper is to present the methodology, the technical and software control solutions that permitted to realize a flexible and accurate QE device with scanning capabilities.
This quantum efficiency setup is indeed highly adaptable and can accommodate different sample sizes and shapes, making it a versatile tool for the study of PMTs with an open door to other photo-sensors in general.
The control of the beam over the sensitive area with an accuracy at the level of $\mu$m can permit high spatial resolution scan and multi channel sensors scans.
The large output power of the Xenon source can provide continuous light with calibrated intensity over six orders of magnitude.This dynamics is matching the range of the power meter and of the picoammeter. Allowing to perform measurements for different sensitivities. 
The setup has been tested and validated by measuring the quantum efficiency of photomultiplier tubes of different brands. The results demonstrate the high accuracy and repeatability of the setup, as well as its ability to map the quantum efficiency over the surface of extended photo-sensors with unprecedented resolution. In summary, this paper presents a novel quantum efficiency setup based on a 2D motorized stage, a wide spectrum xenon lamp, a beam splitter system, and two calibrated photo-diodes for the precise measurement of quantum efficiency of PMTs photo-diodes over the range of 250 to 1100 nm.
\section *{Acknowledgments}
This work was supported by the Italian Ministero dell’Università
e della Ricerca (MUR), PON R\&I program (Project PACK-PIR01 00021).
A. Simonelli has been supported by a grant nr 23392 of the Italian Ministero dell’Università e della Ricerca (MUR), Progetto CIR01 00021 dal titolo: 'Potenziamento Appulo-Campano di KM3- NeT - Rafforzamento del capitale umano' (Avviso n. 2595 del 24 dicembre 2019)

\bibliography{biblio.bib}
\bibliographystyle{unsrt}
\end{document}